\documentclass[12pt, a4paper,oneside,preprint]{article}

\usepackage[margin=1.5cm]{geometry}
\usepackage{amsmath,amssymb,amsfonts,cite,graphicx,sidecap,float,wrapfig,caption,subfig}
\usepackage[utf8]{inputenc}
\usepackage[affil-it]{authblk}
\usepackage{color}

\begin{document}
\renewcommand{\familydefault}{\sfdefault}
\title{A high-fidelity photon gun: \\intensity-squeezed light from a single molecule}
\author{Xiao-Liu Chu, Stephan G\"otzinger, and Vahid Sandoghdar}
\affil{Max Planck Institute for the Science of Light, 91058 Erlangen, Germany}
\affil{Department of Physics, Friedrich Alexander University, 91058 Erlangen, Germany}
\affil{Graduate School in Advanced Optical Technologies (SAOT), 91058 Erlangen, Germany}

\maketitle

\textbf{A two-level atom cannot emit more than one photon at a time. As early as the 1980s, this quantum feature was identified as a gateway to ``single-photon sources", where a regular excitation sequence would create a stream of light particles with photon number fluctuations below the shot noise \cite{Mandel79}. Such an intensity squeezed beam of light would be desirable for a range of applications such as quantum imaging, sensing, enhanced precision measurements and information processing \cite{Lounis05, Polyakov09}. However, experimental realizations of these sources have been hindered by large losses caused by low photon collection efficiencies and photophysical shortcomings. By using a planar metallo-dielectric antenna applied to an organic molecule, we demonstrate the most regular stream of single photons reported to date. Measured intensity fluctuations reveal 2.2\,dB squeezing limited by our detection efficiency, equivalent to 6.2\,dB intensity squeezing right after the antenna. }  

Single-photon sources (SPS) have attracted a great deal of attention in the past two decades. One of the common methods for generating single photons is based on parametric down conversion in nonlinear materials, whereby a pair of photons are produced and separated using spectral, spatial, or polarization filters. Advantages of this approach are its large wavelength tunability and the possibility to herald the detection of one photon by the other one of the pair. However, the emission statistics in this method remains Poissonian, and it cannot produce single photons on demand. 

To exclude the generation of two simultaneous photons, and more importantly, to achieve deterministic production of single photons at a given time, the antibunched radiation of an isolated quantum emitter offers the best choice \cite{Kimble77, Teich88, Kwiat95}. However, so far random losses in the emission, collection, and detection processes have made it impossible to register one single photon at a particular time \cite{Gisin:02}, so that the common terminologies of ``single-photon" source or photon ``gun" used in the literature greatly fall short of their expectations. 

Three sources of loss have to be controlled. First, photophysical losses caused by a non-unity quantum efficiency (e.g. for color centers \cite{Mohtashami:13}) or unforseeable transitions to dark states (e.g. in colloidal quantum dots \cite{Kuno:01}) spoil a deterministic emission. Second, it is a great challenge to collect all the photons, which are usually emitted in a nearly isotropic fashion. Third, the quantum efficiency of the detectors and losses in the optical path limit the final performance. Here, we present a substantial progress in remedying these issues by using single organic molecules with very high quantum efficiency as emitter, optimizing the excitation pulse scheme, careful choice of the optical elements, and employing a new design of metallo-dielectric antennas \cite{Chen:11, Chu:14}. 

\begin{figure}[h!t]
\includegraphics[width = 1\columnwidth]{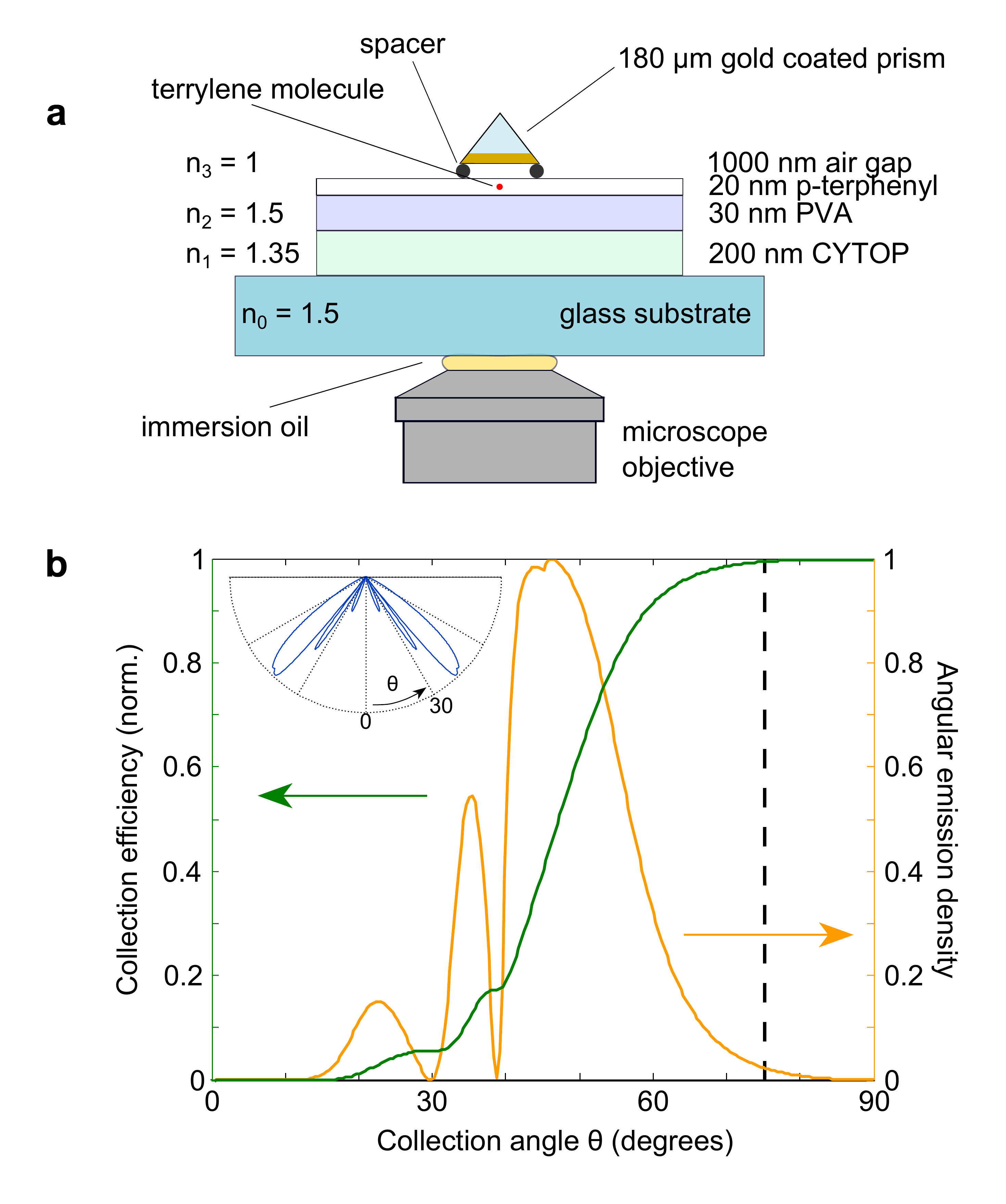}
\caption{\textbf{Antenna design for 99\% collection efficiency from a single molecule.} a, Sketch of the metallo-dielectric antenna. PVA and CYTOP act as quasi-waveguide on top of a cover glass, while a $180\,\mu$m gold coated prism prevents losses to the upper half space. Various elements are not drawn to scale. b, Emitted power density and collection efficiency as a function of the collection angle. The vertical dashed line indicates the maximum collection angle of the microscope objective. Inset: emitted power density in polar coordinates.}
\label{schematics}
\end{figure}

The structure of the new antenna is depicted in Fig. \ref{schematics}a. By taking a standard cover glass as the high index bottom layer, we are able to use a microscope objective that works with a highly transmissive immersion oil. In the particular realization of our current work, we place a 30\,nm thick PVA film (refractive index 1.5) on top of a 200\,nm thick CYTOP layer (refractive index 1.35) to form a quasi-waveguide and subsequently spin coat a 20\,nm layer of p-terphenyl, which is doped with terrylene molecules. These molecules are well known for their photostability at room temperature and fluoresce at a wavelength of about 590\, nm \cite{Pfab04}. As the top metallic mirror we used a gold-coated microprism, which we placed on top of polystyrene beads (diameter $1\,\mu$m) \cite{footnote-chu:16} to avoid the coupling of terrylene fluorescence to surface plasmons and consequent losses \cite{Chen:11}. Using a combination of a planar antenna structure and an objective with a numerical aperture (NA) of 1.46, we can collect more than $99\%$ of the emitted photons (see fig. \ref{schematics}b). The inset in Figure 1b shows the modified radiation pattern of a single oriented dipole inside the metallo-dielectric antenna. 

\begin{figure}[h!t]
\includegraphics[width = 1\columnwidth]{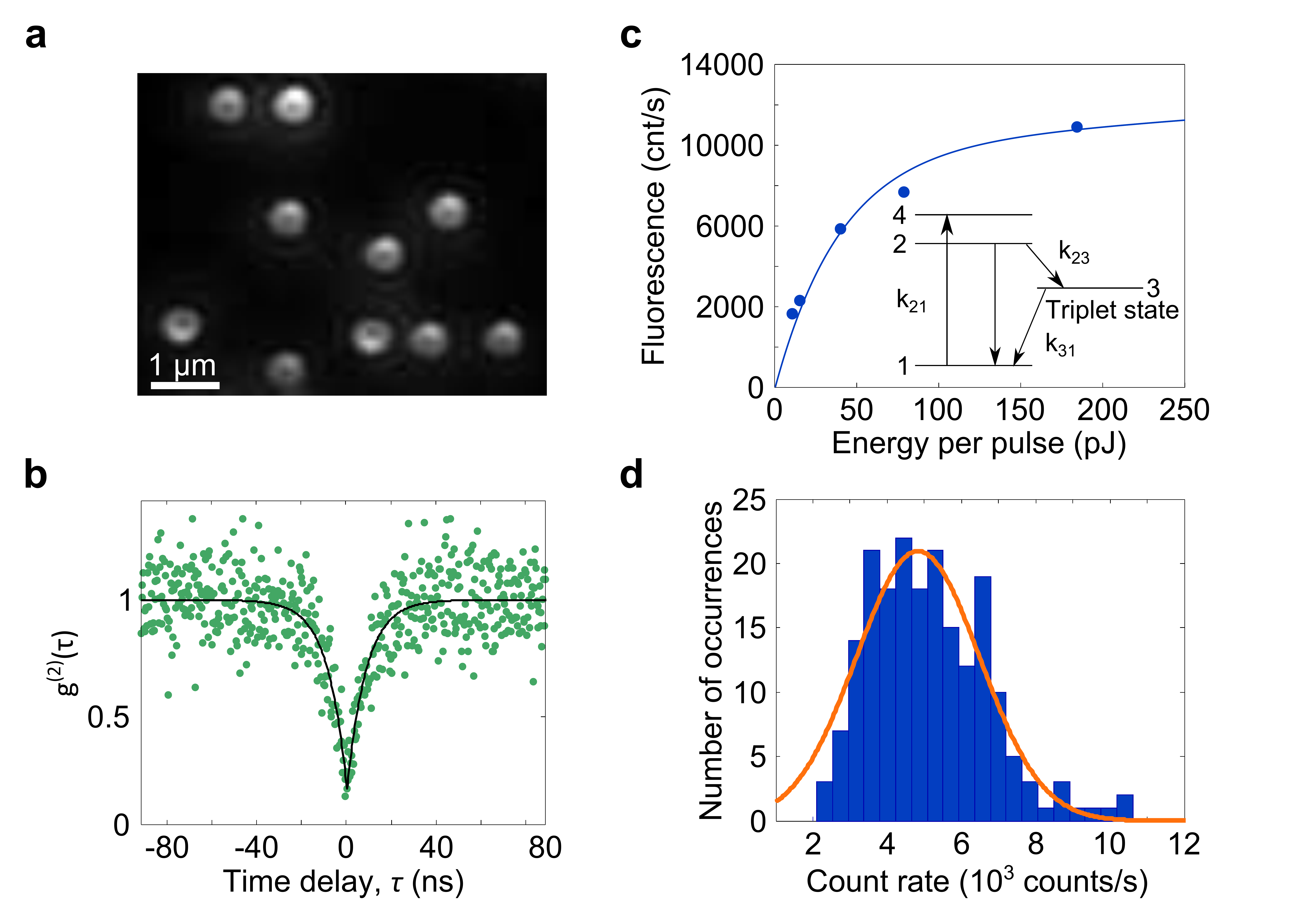}
\caption{\textbf{Characterization of single-molecule single-photon sources.} a, Wide-field image of single molecules in the metallo-dielectic antenna. Individual terrylene molecules are clearly recognized by their characteristic doughnut pattern. b, Measured normalized second-order correlation function of a single molecules under continuous-wave strong excitation. c, Saturation measurement of a single molecule. The repetition rate of the pulsed laser was set to 15\,kHz. Note the high photon detection rate due to the metallo-dielectic antenna.  The inset shows a simplified Jablonski diagram with the relevant transition rates. $k_{21}\approx1\times10^8$, $k_{23}\approx1\times10^4$, $k_{31}\approx6\times10^3$. d, Histogram of the fluorescence count rates obtained from $198$ molecules driven at maximum pulse energy.}
\label{characterization}
\end{figure}

The crystalline p-terphenyl sample is prepared at a low enough terrylene concentration such that these molecules are spatially well separated. Thus, as shown in Fig. \ref{characterization}a, each terrylene molecule can be individually addressed via p-polarized excitation in total internal reflection mode, appearing as a doughnut-shaped spot since its dipole orientation is nearly perpendicular to the interface \cite{Pfab04}. The second-order correlation function shown in Fig \ref{characterization}b shows a value of $g^{(2)}(0)=0.15$ and confirms the isolation of a single molecule. The measurement also indicates that a second photon might accompany the photon emitted by the molecule in about $7.5\%$ of the excitations, consistent with the background fluorescence recorded on the sample next to the molecule. Considering that this background fluorescence can be modeled as a weak Poissonian light source, we can assume that the probability of simultaneous emission of two background photons is $0.3\%$. Since our detector is not number resolving, we measure the background at a given excitation power and account for it in the noise analysis. 

The simplified energy level scheme of terrylene is shown in the inset of Fig. \ref{characterization}c. Considering an expected dark period of about 170 $\mu$s for the triplet state \cite{Lee11}, we chose the excitation repetition rate to be as low as 15\,kHz so that the interval between consecutive pulses is long enough for the molecule to return to the ground state before the arrival of the next excitation pulse. In this fashion, we reduced losses due to intersystem crossing to about $0.01\%$. Figure \,\ref{characterization}c shows the fluorescence signal of a single molecule as a function of the excitation power. The data is fitted with the function \cite{Treussart2002},
\begin{equation}
R(E_{\rm p})=R_{0} \rho(E_{\rm p})+\alpha E_{\rm p},
\end{equation}
where $E_{\rm p}$ represents the pulse energy, $\rho(E_{p})$ is the excited state population right after the excitation pulse, $\alpha E_{\rm p}$ accounts for the background fluorescence, and $R_{0}$ is the maximum attainable photon count rate. Furthermore, the excited state population is given by \cite{Treussart2002}:
\begin{equation}
\rho(E_{\rm p})=\frac{E_{\rm p}/E_{\rm s}}{1+E_{\rm p}/E_{\rm s}}(1-e^{-(\tau_{\rm p}/\tau_{\rm r})[1+(E_{\rm p}/E_{\rm s})]}),
\end{equation}
where $E_{\rm s}$ is the saturation energy, $\tau_{\rm p}$ is the excitation pulse length, and $\tau_{\rm r}$ is the lifetime of the excited state. It turns out that we can excite a molecule with $99\%$ probability at about 200\,pJ, which was the largest available pulse energy in our setup (Fig. \ref{characterization}c).

In addition to dark states, the quantum efficiency of the emitter can severely compromise the performance of a SPS. Figure\,\ref{characterization}d plots a histogram of the emission count rate for nearly 200 molecules upon strong excitation. While the quantum efficiency of terrylene in p-terphenyl is known to reach values beyond 95\% \cite{buchler05}, our measurements reveal a substantial inhomogeneity that leads to a distribution of the emitted powers. This behavior is also common for other solid-state systems such as color centers in diamond \cite{Mohtashami:13}. Thus, to achieve the lowest loss level, we performed our experiments on molecules with a high emission rate. In particular, the saturation measurement displayed in Fig. \ref{characterization}c was obtained on a molecule at an average count rate of 11400 counts/s, which is in very good agreement with the measured overall detection efficiency of 68\%, excitation rate of 15\,kHz, and a background level measured at 1100 counts/s (see Methods). In other words, we detect a photon on demand with a fidelity of about 70\%. 

\begin{figure*}[h!t]
\includegraphics[width = 1\textwidth]{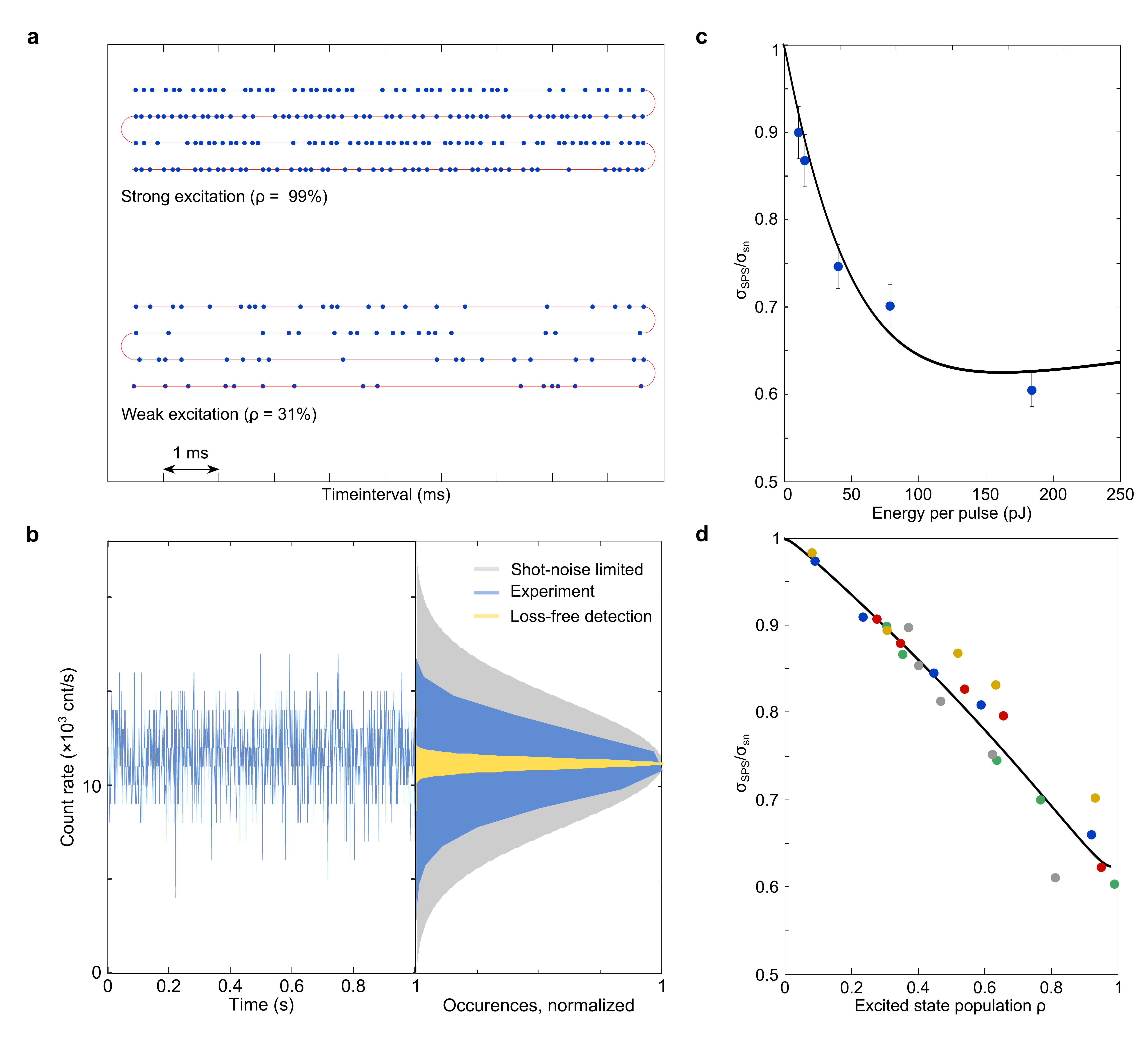}
\caption{\textbf{A regular stream of single photons leads to intensity squeezed light.} a, Detection of single photons emitted by a molecular SPS. At high excitation energies ($\rho=0.99$) the detection events are mostly regular whereas for the same molecule driven in the weak excitation regime ($\rho=0.31$) the events become random. b, Left panel: Intensity trace of a strongly driven single molecule. Right panel: The measured intensity noise over 10\,s (blue) is reduced by $40\%$ compared to a shot-noise limited source of the same intensity. The yellow curve shows the noise distribution of the bare single-photon source including the contribution of background fluorescence, i.e. right after the antenna. c, Normalized intensity noise of a single-molecule single-photon source vs the excitation pulse energy. The positive slope of the theory curve at higher pulse energies originates from linearly increasing background. d, Measured intensity squeezing obtained from five different molecules (each shown in different color) and for various excited state populations $\rho(E_{p})$ deduced from saturation measurements.}
\label{results}
\end{figure*}

To portray the remarkable quality of our SPS, in the upper part of Fig.\,\ref{results}a we present a portion of the individual photon detection events over a period of 40\,ms. A simple visual inspection confirms the high regularity of the photon arrivals. For comparison, in the lower part of the figure we emulate an example of a more typical SPS trace by lowering the excitation probability to 31\% and, thus, lower efficiency. 

For a quantitative demonstration of the SPS performance, we determine its sub-Poissonian nature from the actual detected count rate and its fluctuations ($\sigma_{\rm sps}$). Figure\,\ref{results}b illustrates a time trace of the detected power. The blue distribution on the right panel shows that the intensity noise is reduced by $40\%$ compared to the fluctuations ($\sigma_{\rm sn}$) of a shot-noise limited source of the same intensity depicted in grey. 

In Fig. \ref{results}c, we show $\sigma_{\rm sps}/\sigma_{\rm sn}$ for the same molecule as a function of the excitation pulse energy. The black theoretical curve takes into account the additional noise due to the background and provides a very good fit. In Fig. \ref{results}d, we display similar data for five different molecules plotted against the excited state population. Again, the data can be fitted very well.

The best result of our measurements reaches $\sigma_{\rm sps}/\sigma_{\rm sn}=0.6$, corresponding to an intensity squeezing of 2.2\,dB. The theoretically predicted ratio reads \cite{Lounis05, Fox06}
\begin{equation}
\sigma_{\rm sps}/\sigma_{\rm sn}=\sqrt{1-\zeta\rho}\,,
\end{equation}
yielding a value of $0.62 \pm 0.03$, in very good agreement with the experimental result, including the measured background. Here, $\zeta$ denotes an overall detection efficiency, taking into account the photophysics of the emitter as well as losses in the detection path. The yellow curve presents the intrinsic noise distribution of our SPS once we eliminate the detection losses, which are currently dominated by the efficiency of our single-photon detector at a value of 80\% (see Methods). This would yield $\sigma_{\rm sps}/\sigma_{\rm sn}=0.24$ equivalent to 6.2\,dB intensity squeezing, limited by the background level of our sample. 

A convenient measure for quantifying the sub-Poissonian nature of a SPS is the so-called Mandel parameter given by $Q_{\rm D}=[g^{(2)}(0)-1]M\zeta\rho$, where $M$ is the average number of photons per excitation pulse \cite{Lounis:00}. For an ideal SPS with unity detection efficiency, $Q_{\rm D}=-1$, but $Q_{\rm D}$ is limited by the total efficiency $\zeta$ in any realistic experiment. The Mandel parameter for the overall detection efficiency of our experiment is -0.64 while the $Q_{\rm D}$ associated with our planar antenna amounts to -0.93. To our knowledge, these amount to the lowest Mandel parameters reported to date for a SPS.

Today, we are still faced with several challenges before SPSs can become available in a manner similar to lasers, i.e. at different wavelengths, bandwidths, or output power levels. To address some of these issues, various materials have been explored, including atoms \cite{Kuhn02, Mckeever04}, molecules\cite{Basche:92b, Lounis:97, Lounis:00}, quantum dots \cite{Imamoglu94, Kim99}, color centers \cite{Kurtsiefer00}, ions in traps \cite{Diedrich87, Keller2004} and ions in a solid \cite{Kolesov12optical, Eichhammer:15}. In addition, different strategies for efficient collection of single photons have also been pursued, encompassing cavity-based concepts \cite{Strauf07, Michler00} as well as wave guiding platforms \cite{Liebermeister14, Claudon10, Arcari14}. The previous efforts do not reach a very high total efficiency and often require sophisticated fabrication or cryogenic operation. In this work, we have presented a straightforward and facile strategy for the generation of a photon on demand and have demonstrated its \textit{detection} at a high fidelity of about 70\%. This performance already surpasses the threshold reported for successful implementation of linear optical quantum computing \cite{Varnava:08}, but forseeable progress in detector technology should make it possible to still improve it by a considerable amount.

The sub-shot-noise light beam generated by a single quantum emitter is also very promising as a new intensity standard and can be used to calibrate photodetectors and counters at low light levels. By employing a low-loss mode conversion \cite{Labroille14}, we plan to couple our device to single mode fibers, delivering convenient sources of nonclassical light for various applications such as quantum key distribution \cite{Gisin:02, Scarani09, Waks02b}, ultra-sensitive detection, imaging, and spectroscopy \cite{Brida10}. Our approach based on planar antennas is also applicable to other solid-state quantum emitters with high quantum efficiency.  

\smallskip

\textbf{Acknowledgements}
We thank Xue-Wen Chen for help with the antenna design and simulation. This project was supported by the European Union (ERC Advanced Grant SINGLEION and the SIQUTE project of the European Metrology Research Program (EMRP)), an Alexander von Humboldt professorship, and the Max Planck Society.

\newpage
\textbf{Methods}
\\
\textbf{Sample fabrication.}
Cover glasses ($170\,\mu$m thickness) were cleaned in an oxygen plasma followed by treatments in methanol, ethanol and isopropanol to remove any organic contamination. Next, a 200\,nm layer of CYTOP was spin coated onto the cleaned cover glass and a 4 nm layer of $\rm TiO_2$ was deposited on top. A $30\,nm$ layer of polyvinylalcohol (PVA) was then used to protect the CYTOP (AGC Chemicals europe ltd.). A solution of toluene containing terrylene and p-terphenyl molecules was then spin coated \cite{Pfab04}. Finally, we placed a $180\,\mu$m gold coated prism onto the sample. However, to allow for sufficient space between the gold layer and the molecule (i.e., avoid coupling to surface plasmons), polystyrene beads were distributed on the p-terphenyl layer first. We confirmed a typical gap of 800-1000\,nm by back focal plane imaging of single molecules.  

\textbf{Experimental setup.}
To excite the terrylene molecules, we used a pulsed laser (time-bandwidth Cheetah-X-SHG) with a pulse width of 13\,ps at a wavelength of 532 nm. After the laser we inserted a pulse picker in order to reduce the repetition rate of the laser to 15\,kHz. The light was focused to an area of about $1\,\mu$m$\times 1\,\mu$m via total internal reflection through the microscope objective. By setting the incident light to be p-polarized, we ensured efficient excitation of the vertically oriented molecular dipole moments. Photons emitted by the molecules were collected by the same microscope objective, separated from the excitation light by a long-pass filter (Semrock, 538 nm) and detected by a sensitive CCD camera, a single-photon counter or a Hanbury-Brown and Twiss photon correlator.

In order to minimize losses, we optimized the setup for a high photon detection efficiency. The photons were collected by an immersion objective (Zeiss alpha Plan-Apochromat 100x/1.46, collection angle 74$^{\circ}$degrees) with a transmission of $90\pm 2\%$. The losses of all the optical components, including mirrors, lenses and filters added up to $5\pm 2\%$. The single photon detector (Laser 2000) had a quantum efficiency of $80\pm 4\%$ at 605\,nm. This allows us to determine the total detection efficiency $\zeta$ to be as high as $68\pm 3\%$.   

\textbf{Noise analysis.}
We model all losses including the detector inefficiency by a single beam splitter with splitting ratio $T/(1-T)$, where the transmission $T$ is equivalent to the total probability of detecting a photon upon excitation of the single molecule \cite{Fox06}. The beam splitter process occurs at the single photon level randomly with a weighted probability of $T$ and $(1-T)$, respectively, according to random sampling. The randomness induced by the beam splitter increases the noise in the transmitted single photon stream towards Poissonian statistics. In order to study the intensity fluctuations of our SPS, we record the photon flux over a period of 10\,s and divide the obtained data into time bins of length 1\,ms. By looking at the variation of the number of detected photons in the time bins, we obtain the intensity noise of the SPS. For the theoretical analysis we model our source as a single emitter where we detect the emitted photons with a certain total detection efficiency and an uncorrelated linear background. The standard deviation is therefore given by:
\[
\sigma_{\rm sps}(E_{p})=\sqrt{\sigma_{\rm mol}^{2}(E_{p})+\sigma_{\rm bg}^{2}(E_{p})},
\]
where $\sigma_{\rm mol}$ and $\sigma_{\rm bg}$ is the noise originating from variations in photons emitted by the single molecule and the Poissonian noise of the background, respectively. Note that the intensity noise of the emitter $\sigma_{\rm mol}(E_{p})=\sqrt{1-\zeta\rho(E_{p})} \,\sigma_{sn}$ contains all loss channels including the detector efficiency and excitation probability.


\begin{thebibliography}{10}
\bibitem{footnote-chu:16}
{In some cases a slight roughness of the p-terphenyl film was sufficient as a
  spacer.}
\end{thebibliography}
\end{document}